\documentclass[rmp,twocolumn,amsmath,amssymb,amsfonts,showpacs,preprintnumbers]{revtex4}

\usepackage{graphicx}

\usepackage{MnSymbol}

\input xy
\xyoption{all}

\begin{document}

\title{Heisenberg--Weyl algebra revisited:  Combinatorics of words and paths}

\author{P. Blasiak}
\email{Pawel.Blasiak@ifj.edu.pl}

\homepage{http://www.ifj.edu.pl/~blasiak}

\author{A. Horzela}
\email{Andrzej.Horzela@ifj.edu.pl}

\affiliation{H. Niewodnicza\'nski Institute of
Nuclear Physics, Polish Academy of Sciences\newline
ul. Radzikowskiego 152,  PL 31342 Krak\'ow, Poland}

\author{G. H. E. Duchamp}
\email{ghed@lipn-univ.paris13.fr}
\affiliation{LIPN, UMR 7030, Universit\'e Paris 13 - CNRS\\
99, av. J-B Cl\'ement, F-93430 Villetaneuse, France}

\author{K. A. Penson}
\email{penson@lptmc.jussieu.fr}

\author{A. I. Solomon}
\email{a.i.solomon@open.ac.uk}

\altaffiliation[\newline Also at ]{The Open University, Physics and Astronomy Department,
Milton Keynes MK7 6AA, United Kingdom}

\affiliation{Laboratoire de Physique Th\'eorique de la Mati\`{e}re Condens\'{e}e\\
Universit\'e Pierre et Marie Curie, CNRS UMR 7600\\
Tour 24 - 2i\`{e}me \'et., 4 pl. Jussieu, F 75252 Paris Cedex 05, France}

\preprint{\emph{J. Phys. A: Math. Theor. }\textbf{41} 415204 (2008)}

\begin{abstract}
The Heisenberg--Weyl algebra, which underlies virtually all physical representations of Quantum Theory, is considered from the combinatorial point of view. We provide a concrete model of the algebra in terms of paths on a lattice with some decomposition rules. We also discuss the rook problem on the associated Ferrers board; this is related to the calculus in the normally ordered basis. From this starting point we explore a combinatorial underpinning of the Heisenberg--Weyl algebra, which offers novel perspectives, methods and applications.
\end{abstract}

                           
\maketitle

\section{Introduction}

$|\_$From a modern viewpoint the formalism and structure of Quantum Theory is founded on the theory of Hilbert space \cite{Isham,Peres}. The physical content of the theory consists of   representing physical quantities as operators which satisfy some algebraic relations. Virtually all correspondence schemes come endowed with the Heisenberg--Weyl algebra structure, be it the canonical quantization scheme, the occupation number representation in quantum mechanics, or the second quantization formalism of quantum field theory. This derives from the analogy with classical mechanics whose Poissonian structure is reflected in the commutator of position and momentum observables \cite{Dirac}. Ubiquitous and profound, the Heisenberg--Weyl algebra has become the hallmark of non-commutativity in Quantum Theory.



An exemplary model of the Heisenberg--Weyl algebra involves combinations of derivative $D$, multiplication $X$ and identity $I$ operators acting on the space of polynomials.
Physical examples are the position $\hat{x}$ and momentum $\hat{p}$ operators in the space of square integrable functions, or the annihilation $a$ and creation $a^\dag$ operators in  Fock space. Here, without loss of generality, we conform to the notation $\{a,a^\dag\}$ for the generators of the  (associative) algebra\footnote{We do not attach much weight to this particular realization, however, as we shall study algebraic properties only, for which the underlying Fock space plays no role. Our considerations hold true for any representation of the Heisenberg--Weyl algebra.}, satisfying
\begin{eqnarray}\label{aa}
aa^\dag=a^\dag a+I,
\end{eqnarray}
where $I$ is the multiplicative identity.
We shall be interested in combinatorial aspects of this relation and discuss one of the ensuing models of the Heisenberg--Weyl algebra.

Indeed, the combinatorial properties  of Eq.(\ref{aa}) were  recognized early and successfully applied to the domain of algebraic enumeration, principally  concerning the action of the operators $X$ and $D$ on generating series. From this point of view these operators are auxiliary constructions facilitating enumeration of discrete structures \cite{Riordan,Stanley,Comtet,Wilf,Flajolet,Bergeron,Bryant}.

However, in this note we adopt  another approach, which leads towards a combinatorial model of the algebra itself.  Starting from the definition of the Heisenberg--Weyl algebra as the algebra of words in $a$ and $a^\dag$ supplemented by the relation of Eq.~(\ref{aa}) we will recast it in the language of paths on a lattice with some decomposition rules. We shall also consider a convenient choice of  basis, here taken to be normally ordered monomials, which permits  a direct link to the combinatorics of words and the related algebra of paths. In this way algebraic problems may be expressed in the  more concrete form of the decomposition and enumeration of paths. For illustration we consider the normal ordering procedure which reduces to the familiar rook problem on the Ferrers board and then derive the structure constants of the algebra by a simple combinatorial argument.

\section{Heisenberg--Weyl algebra}

In this note we consider an \emph{algebra} $\mathcal{A}$ to be a linear vector space over a field $\mathbb{K}$ with bilinear multiplication law
\begin{eqnarray}
\mathcal{A}\times\mathcal{A}\ni(b,c)\longrightarrow b\,c\in\mathcal{A}
\end{eqnarray}
which is associative and possesses a  unit element $I$. More precisely, it is called an \emph{associative algebra with unit} as distinct from an algebraic structure lacking associativity or a unit (\emph{e.g.} Lie algebra). A \emph{basis} of an algebra is a basis for its vector space structure. Each basis $(b_i)_{i\in \Lambda}$ defines a unique family $\gamma_{ij}^k\in\mathbb{K}$ such that for every ordered pair $(i,j)\in \Lambda\times \Lambda$ the set of $k\in \Lambda$ such that $\gamma_{ij}^k\neq 0$ is finite and
\begin{eqnarray}\label{StrConst}
b_i\,b_j=\sum_{k\in \Lambda}\gamma_{ij}^k\,b_k\,.
\end{eqnarray}
The $\gamma_{ij}^k$ are called the \emph{structure constants} of the algebra $\mathcal{A}$ with respect to the basis $(b_i)_{i\in \Lambda}$, from which the multiplication law can be uniquely recovered.

The \emph{Heisenberg--Weyl algebra}, denoted by $\mathcal{H}$, is the algebra generated by $a$, $a^\dag$, satisfying the relation of Eq.~(\ref{aa}).
Elements of the algebra $A\in\mathcal{H}$ are linear combinations of finite products of $a$ and $a^\dag$ of the form
\begin{eqnarray}\label{Aaaaaa}
A=\sum_{\textbf{r},\textbf{s}}\ \ \alpha_{\textbf{r},\textbf{s}}\ \ a^{\dag\,r_1} a^{s_1}a^{\dag\,r_2} a^{s_2}\,...\ a^{\dag\,r_k} a^{s_k},
\end{eqnarray}
where $\textbf{r}=(r_1,...,r_k)$ and $\textbf{s}=(s_1,...,s_k)$ are nonnegative integer multi-indexes (with the convention $a^0=a^{\dag\,0}=I$). This representation is ambiguous, however, due to the commutation relation Eq.~(\ref{aa})
which yields different representations of the same element of the algebra, \emph{e.g.} $aa^\dag=a^\dag a +I$.
The problem can be resolved by fixing the preferred order of the generators $a$ and $a^\dag$.
Conventionally, it is done by choosing the \emph{normally ordered} form in which
all annihilators stand to the right of creators.
In this case, each element of the algebra $\mathcal{H}$ is uniquely written in the normally ordered form as
\begin{eqnarray}\label{A}
A=\sum_{r,s}\beta_{rs}\ a^{\dag\,r} a^s\,.
\end{eqnarray}
Hence the normally ordered monomials $a^{\dag\,r}a^s$ constitute a natural basis for the Heisenberg--Weyl algebra
\begin{eqnarray}\label{Basis}
\textsc{Basis\ of\ } \mathcal{H}\ : \ \ \ \ b_{\textit{(r,s)}}=a^{\dag\,r}a^s\,,
\end{eqnarray}
indexed by pairs of integers $r,s=0,1,2,...$, and Eq.~(\ref{A}) is the expansion of the element $A$ in this basis. We should note that the normally ordered representation of the elements of the algebra suggests itself as the simplest one \cite{CahillGlauber}. It is important and commonly used in practical applications in quantum optics \cite{Glauber,Schleich,Klauder} or quantum field theory \cite{BjorkenDrell,Mattuck}. Working in this basis entails the reshuffling of $a$ and $a^\dag$ to the normally ordered form, which in general is a nontrivial task \cite{AmJPhys}. This brings up the issue of efficient calculation methods and intuitive schemes providing insight into the ordering procedure itself. Below, we provide a combinatorial model for the Heisenberg--Weyl algebra and then propose a resolution of the problem from this starting-point.

\section{Combinatorics of the Heisenberg--Weyl algebra}

\subsection{Words, paths and rook problems}\label{Paths}

We start by showing that each word in a two-letter alphabet can be uniquely encoded as a \emph{staircase path} on a plane rectangular lattice. This observation will lead us to considering the associated Ferrers board and the rook problem. Rather than giving a formal construction, we shall illustrate it by an example from which the general scheme can be straightforwardly recovered.

\begin{figure}
\begin{center}
\resizebox{\columnwidth}{!}{\includegraphics{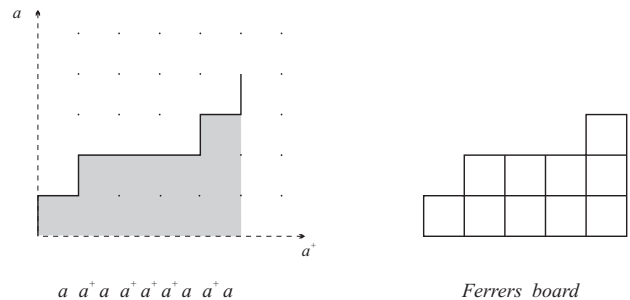}}
\end{center}
\caption{\label{Fig1} Path defined by the word $\texttt{w}$ and the associated Ferrers board $\mathcal{B}_\texttt{w}$.}
\end{figure}

Suppose we consider a word, say
\begin{eqnarray}\label{w}
\texttt{w}= a\,a^\dag a\,a^\dag a^\dag a^\dag a\,a^\dag a
\end{eqnarray}
to which we assign a staircase path on a rectangular lattice. Starting from the point $(0,0)$, it is constructed by reading the word $\texttt{w}$ from the left and drawing a line to the right if the letter is $a^\dag$ and up if the letter is $a$, as shown in Fig.~\ref{Fig1} on the left.
We observe that this scheme provides a unique encoding of words.

\begin{figure*}
\begin{center}
\resizebox{1.4\columnwidth}{!}{\includegraphics{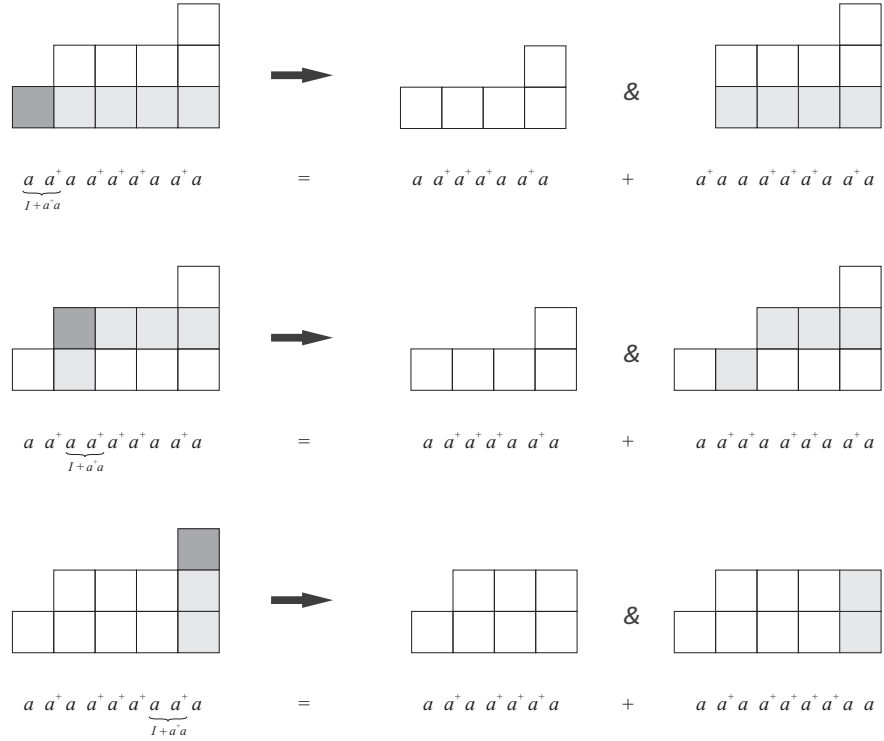}}
\end{center}
\caption{\label{Fig2} Three possible decompositions of the Ferrers board $\mathcal{B}_\texttt{w}$ and the corresponding reduction of the word $\texttt{w}$.}
\end{figure*}
With each path (or word) one can associate a {\em Ferrers board} $\mathcal{B}_\texttt{w}$ by retaining rectangular cells below the path, see Fig.~\ref{Fig1}. Note that this is a many-one procedure as paths differing by a horizontal line at the beginning and a vertical line at the end yield the same board \cite{Stanley,Bryant}.

The \emph{rook problem} for the given board $\mathcal{B}$ consists in enumerating non--capturing arrangements of $k$ rooks on the board which defines a finite sequence $r_\mathcal{B}(k)$, $k=0,1,2,...$, called the \textit{rook numbers}.
In our example one has
\begin{eqnarray}\label{numbers}
r_\mathcal{B}(k)=1,10,23,9,0,0,...\ \ \ \ \ k=0,1,2,3,...\,.
\end{eqnarray}
A rook sequence can be conventionally encoded in a \textit{rook polynomial} defined by
\begin{eqnarray}
R_\mathcal{B}(x)=\sum_{k\geq 0} r_\mathcal{B}(k)\,x^k.
\end{eqnarray}
It is straightforward to show that these polynomials satisfy the recursion \cite{Riordan,Stanley,Bryant}
\begin{eqnarray}\label{RookRec}
&&\begin{array}{l}
R_\mathcal{B}(x)=R_\mathcal{B'}(x)+ x\,R_\mathcal{B''}(x),\\
R_{\text{\O}}(x)=1,
\end{array}
\end{eqnarray}
obtained by choosing a cell which forms a step in the diagram $\mathcal{B}$. This \emph{step-forming} cell has no neighboring cell to its left or above it. We now consider two cases: a rook is placed on the cell or not. This reduces the problem to the boards $\mathcal{B''}$ (with the row and column in which the cell is placed removed) and $\mathcal{B'}$ (with the chosen cell removed only). We note that there are many possible choices of such a cell, and these give different decompositions of the board yielding various recursive patterns, see Fig.~\ref{Fig2}.

\subsection{Normal ordering procedure}\label{NormalOrder}

It has been shown \cite{Navon,QTS3,Varvak} that the normal ordering of a word $\texttt{w}$ in $a$ and $a^\dag$ satisfying Eq.~(\ref{aa}) reduces to the rook problem on the associated Ferrers board $\mathcal{B}_\texttt{w}$. Namely,
\begin{eqnarray}\label{NormalRook}
\texttt{w}=\sum_{k\geq 0} r_{\mathcal{B}_\texttt{w}}(k)\,\texttt{w}^{\,\textsl{(k)}}
\end{eqnarray}
where $\texttt{w}^{\,\textsl{(k)}}$ are  normally ordered monomials $a^{\dag\,r}a^s$ obtained from $\texttt{w}$ by crossing out $k$ pairs of $a$ and $a^\dag$ and then reshuffling the rest as if they were commuting variables (called the double dot operation used in quantum field theory). For example, for a word in Eq.~(\ref{w}) we have $\texttt{w}^{\,\textsl{(0)}}=a^{\dag\,5}a^4$,  $\texttt{w}^{\,\textsl{(1)}}=a^{\dag\,4}a^3$, $\texttt{w}^{\,\textsl{(2)}}=a^{\dag\,3}a^2$, \emph{etc.}, and hence its normally ordered form reads (see Eq.~(\ref{numbers}))
\begin{eqnarray}
\texttt{w}=a^{\dag\,5}a^4+10\,a^{\dag\,4}a^3+23\,a^{\dag\,3}a^2+9\,a^{\dag\,2}a\ .
\end{eqnarray}

A rigorous proof of Eq.~(\ref{NormalRook}) relies on the observation that each word can be reduced to the sum of two simpler ones by choosing the places in which $a$ precedes $a^\dag$ (which correspond to step-forming cells of the previous section) and reshuffling them according to Eq.~(\ref{aa}), \textit{i.e.} $aa^\dag\rightarrow I+a^\dag a$. For example, for a word of Eq.~(\ref{w}) there are three choices
which exactly correspond to possible decompositions of the associated Ferrers diagram $\mathcal{B}_\texttt{w}$ in Fig.~\ref{Fig2}. We note that although there are various possible decomposition schemes it can be shown that the result is unique.

In short, the normal ordering of a word reduces  to the enumeration of possible non-capturing rook arrangements on the associated Ferrers board. The problem can be systematically handled by successive decompositions of the board. Moreover, one can devise simple algorithms based on the recursive rule given in Eq.~(\ref{RookRec}). The methods described in this paper may be extended to the q-deformed case, see \emph{e.g.} \cite{Duchamp1995,Varvak}.

\subsection{Combinatorial realization of the Heisenberg--Weyl algebra}

We observed in Sect.~\ref{Paths} that each word in two letters, here taken as $a$ and $a^\dag$, can be encoded as a path, see Fig.~\ref{Fig1}. This establishes an isomorphism between the algebra $\mathcal{W}$ of words in two letters, and the algebra $\mathcal{P}$ of paths. In $\mathcal{W}$ multiplication is given by simple concatenation of words with the unit being the void word, while in $\mathcal{P}$ multiplication is given by concatenation of paths. In both cases we shall indicate the unit, which is the void word or path respectively, by the symbol $\text{\O}$. Both algebras are free. The Heisenberg--Weyl algebra arises by imposing on $\mathcal{W}$ the relation of Eq.~(\ref{aa}), \textit{i.e.} $\mathcal{H}=\mathcal{W}/_{\{aa^\dag=a^\dag a+ I\}}$. In $\mathcal{P}$ this relation takes the symbolic form
\begin{eqnarray}\label{RelPath}
\lefthalfcap\,=\,\righthalfcup\ +\ \fivedots,
\end{eqnarray}
by which we mean that a given staircase path is equivalent to the sum of two staircase paths obtained by \\
(\emph{i}) replacing an \emph{upper-left-hand} corner ($\lefthalfcap$) by a \emph{lower-right-hand} corner ($\righthalfcup$), and\\
(\emph{ii}) removing the row and column which intersect in the given cell ($\fivedots$).\\
Note that this reduction is exactly equivalent to the decomposition of the associated Ferrers board induced by the rook problem. In fact, any path can be uniquely decomposed into a finite sum of paths without steps (\emph{i.e.} paths pertaining to monomials $a^{\dag\,r}a^s$). The latter constitute the basis in $\mathcal{P}/_{\{\lefthalfcap=\righthalfcup\,+\,\fivedots\}}$ corresponding to the normally ordered basis in $\mathcal{H}$.

In this way, we obtain the combinatorial model of the Heisenberg--Weyl algebra as the algebra of paths with the relation Eq.~(\ref{RelPath}), \textit{i.e.}
\begin{eqnarray}\label{HP}
\mathcal{H}=\mathcal{W}/_{\{aa^\dag=a^\dag a+ I\}}\cong\mathcal{P}/_{\{\lefthalfcap=\righthalfcup\,+\, \fivedots\}}.
\end{eqnarray}
With this equivalence the algebraic structure of $\mathcal{H}$ and its calculus is recast in combinatorial terms, thus providing  different perspectives and making accessible intuitive combinatorial arguments.

\subsection{Example: Structure constants}

\begin{figure}
\begin{center}
\resizebox{\columnwidth}{!}{\includegraphics{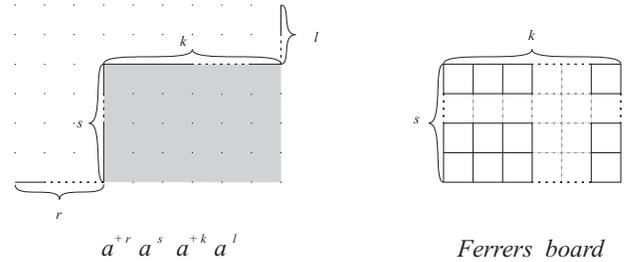}}
\end{center}
\caption{\label{Fig3} Path defined by the word $\texttt{w}=a^{\dag\,r}a^s\, a^{\dag\,k}a^l$ and the associated Ferrers board.}
\end{figure}

To illustrate  the above equivalence we show how to calculate the structure constants Eq.~(\ref{StrConst}) of the Heisenberg--Weyl algebra by simple combinatorial enumeration.
In the form suited for our purposes here we multiply two elements of the basis of Eq.~(\ref{Basis}), say $b_{\textit{(r,s)}}$ and $b_{\textit{(k,l)}}$, defining the structure constants $\gamma_{\textit{(r,s)}\textit{(k,l)}}^{\textit{(p,q)}}$ in the form 
\begin{eqnarray}\label{StrConstHW}
b_{\textit{(r,s)}}\,b_{\textit{(k,l)}}=\sum_{p,q}\gamma_{\textit{(r,s)}\textit{(k,l)}}^{\textit{(p,q)}}\ b_{\textit{(p,q)}}\,.
\end{eqnarray}
Expansion in this basis of the product $b_{\textit{(r,s)}}\,b_{\textit{(k,l)}}$  essentially comes from   the normal ordering of the word
\begin{eqnarray}
\texttt{w}=a^{\dag\,r}a^s\, a^{\dag\,k}a^l\ .
\end{eqnarray}
Following the scheme of Section~\ref{Paths} we draw the associated path and subsequently read off the Ferrers board $\mathcal{B}_\texttt{w}$ which has a simple rectangular form, see Fig.~\ref{Fig3}.
Enumeration of non--capturing rook arrangements on the board is a direct result of the observations:\\ (\emph{1}) the maximum number of rooks on the board is $\text{min}\,\{s,k\}$, and \\
(\emph{2}) there are as many possible arrangements of $i$ rooks as unordered choices of $i$ columns and rows in the board $\mathcal{B}_\texttt{w}$.\\ 
Hence the rook numbers are
\begin{eqnarray}
r_{\mathcal{B}_\texttt{w}}(i)=i!{s\choose i}{k\choose i}\ ,
\end{eqnarray}
for $i=0\ ...\ \text{min}\,\{s,k\}$ and zero otherwise. Consequently, Eq.~(\ref{NormalRook}) gives the normally ordered form of the word $\texttt{w}$
\begin{eqnarray}\label{AAA}
a^{\dag\,r}a^s\, a^{\dag\,k}a^l=\!\sum_{i=0}^{\text{min}\{k,s\}}\!i!{s\choose i}{k\choose i}\,a^{\dag\,{r+k-i}}a^{s+l-i}\ .
\end{eqnarray}
Finally, from Eq.~(\ref{StrConstHW}) one may read off the structure constants of the algebra $\mathcal{H}$ in the normally ordered basis. For fixed $\textit{(r,s)}$ and $\textit{(k,l)}$ the only non-vanishing $\gamma$s are
\begin{eqnarray}\label{gammas}
\gamma_{\textit{(r,s)}\textit{(k,l)}}^{\textit{(r+k-i,s+l-i)}}=i!{s\choose i}{k\choose i}\ \ \ \ \text{for}\ \ i=0\ ...\ \text{min}\,\{k,s\}\ .
\end{eqnarray}
Note that the right hand side of Eq.~(\ref{gammas}) is, in fact,independent of $r$ and $l$ since the outer elements are not included in the commutation.

\section{Summary}

We considered the Heisenberg--Weyl algebra $\mathcal{H}$ starting from the (two generator) free algebra of words $\mathcal{W}$ and imposing the relation of Eq.~(\ref{aa}), \textit{i.e.}

\begin{eqnarray}\nonumber
\xymatrix{
\ \ \ \ \ \ \ \mathcal{W}\ \ \ar[rr]^{\ \ \ \ [a,a^\dag]=I}&&\ \ \mathcal{H}
}
\end{eqnarray}
We showed that words can be uniquely encoded as  staircase paths on a plane, thus providing a realization of $\mathcal{W}$ as a combinatorial algebra of paths $\mathcal{P}$. This allowed us to construct a model of $\mathcal{H}$ in terms of paths with the decomposition rule of Eq.~(\ref{RelPath}) in $\mathcal{P}$ which reflects the defining relation of Eq.~(\ref{aa}) in $\mathcal{W}$. The following diagram illustrates the whole scheme
\begin{eqnarray}\nonumber
\xymatrix{
\ \ \ \ \ \ \ \ 	\mathcal{W}\ \ \ \ \ \ \ \ \ar@{<->}[rr]^{\textstyle{\sim}}\ar[d]&&\ \ \ \ \ \ \ \ \mathcal{P}\ar[d]\ \ \ \ \ \ \ \ \\
\ \ \mathcal{W}/_{aa^\dag=a^\dag a+ I}\ \ \ar@{<->}[rr]^{\textstyle{\sim}}&&\ \ \mathcal{P}/_{\{\lefthalfcap=\righthalfcup\,+\, \fivedots\}}\ \
	}
\end{eqnarray}

We further looked at the normally ordered basis in $\mathcal{H}$ and the corresponding basis in $\mathcal{P}/_{\{\lefthalfcap=\righthalfcup\,+\, \fivedots\}}$.
We pointed out that the decomposition rule of Eq.~(\ref{RelPath}), reducing the number of steps in a path, is closely related to the familiar rook problem on the associated Ferrers board. This permits a  graphical illustration of the calculus in this basis and, more generally, of the normal ordering problem.

In this note we have advocated a combinatorial approach to the Heisenberg--Weyl algebra by showing that it can be conceived as having a purely combinatorial origin. This gives a new perspective on  the whole algebraic framework which is  easily amenable  to the sophisticated methods of discrete mathematics. Finally, we should mention other models of the algebra based on various discrete structures such as set partitions, graphs or urn models \cite{Mendez,Graphs,Urns} as well as results deriving from combinatorial methodology, see \cite{AmJPhys} and references therein.

\section*{Acknowledgments}
We thank Hayat Cheballah for useful discussions.
One of us (P.B.) acknowledges his appreciation of the warm hospitality
and support of the Laboratoire d'Informatique de l'Universit\'e Paris-Nord in Villetaneuse
where most of his research was carried out.
P.B and A.H. wish to acknowledge support from the
Polish Ministry of Science and Higher Education
under Grants No. N202 061434 and N202 107 32/2832.


\end{document}